\documentclass{webofc}
\usepackage[varg]{txfonts}   
\usepackage{hyperref}
\usepackage{xcolor}
\begin{document}
\title{Towards the automation of Monte Carlo simulation on GPU
for particle physics processes}
%
%

\author{\firstname{Stefano} \lastname{Carrazza}\inst{1,2,3}\fnsep\thanks{\email{stefano.carrazza@unimi.it}} \and
        \firstname{Juan} \lastname{Cruz-Martinez}\inst{1}\fnsep\thanks{\email{juan.cruz@mi.infn.it}} \and
        \firstname{Marco} \lastname{Rossi}\inst{1,2}\fnsep\thanks{\email{marco.rossi@cern.ch}} \and
        \firstname{Marco} \lastname{Zaro}\inst{1}\fnsep\thanks{\email{marco.zaro@mi.infn.it}}
}

\institute{Dipartimento di Fisica, Universit\`a degli Studi di Milano and INFN Sezione di Milano, Milan, Italy
\and
           CERN, Theoretical Physics Department and OpenLab, CH-1211 Geneva 23, Switzerland.
\and
           Quantum Research Centre, Technology Innovation Institute, Abu Dhabi, UAE.
          }

\abstract{
In this proceedings we present {\tt MadFlow}, a new framework for the automation
of Monte Carlo (MC) simulation on graphics processing units (GPU) for particle
physics processes.
In order to automate MC simulation for a generic number of processes, we design
a program which provides to the user the possibility to simulate custom
processes through the MadGraph5\_aMC@NLO framework. The pipeline includes a first
stage where the analytic expressions for matrix elements and phase space are
generated and exported in a GPU-like format.
The simulation is then performed using the VegasFlow and PDFFlow libraries which
deploy automatically the full simulation on systems with different hardware
acceleration capabilities, such as multi-threading CPU, single-GPU and multi-GPU
setups.
We show some preliminary results for leading-order simulations on different
hardware configurations.}
\maketitle

\section{Introduction and motivation}

In the last years we have observed a quickly growing interest on the development of new code frameworks
based on hardware accelerators, in particular graphics processing units (GPU),
for the improvement of performance and efficiency in scientific and industrial
problems.

There are several examples in High Energy Physics (HEP) applications that can
benefit from a systematic implementation (or conversion) of existing codes and
algorithms on GPU. Some examples have already been successfully published, such
as deep learning applications in experimental physics~\cite{Albertsson:2018maf},
where spectacular performance improvements were obtained thanks to the
employment of GPUs.

The HEP community has been very active in the field during the last years, for
example providing computational tools for experimental setups. On the other
hand, we still observe a growing trend towards the increase of computational
time required to solve complex problems~\cite{Hoche:2013zja, Niehues:2018was}
in particular for Monte Carlo (MC) simulation of particle physics processes and
observables.
In fact, public MC simulation libraries rely, almost exclusively,
on CPU technology~\cite{Gleisberg:2008ta, Alwall:2014hca, Frederix:2018nkq, Campbell:2019dru}.

In order to write a full parton-level MC competitive by any measure with existing tools
at least four ingredients need to be produced: an integrator library able to parallelize
over the number of events; a GPU-capable PDF interpolation tool;
an efficient phase space generator, which should not only generate valid phase space points on GPU
but also apply any fiducial cuts and finally the matrix element squared for the target processes
are required.

In the last year, some of us have developed tools that provide the ground basis
for the implementation of an automatic Monte Carlo simulation tool for
HEP addressing some of the aforementioned issues:
VegasFlow~\cite{Carrazza:2020rdn,Carrazza:2020loc} and
PDFFlow~\cite{Carrazza:2020qwu,Rossi:2020sbh}. VegasFlow is a new software for
fast evaluation of high dimensional integrals based on Monte Carlo integration
techniques designed for platforms with hardware accelerators. It enables
developers to delegate all complicated aspects of hardware or platform
implementation to the library, so they can focus on the problem at hand. PDFFlow
is a library which provides fast evaluation of parton distribution functions
(PDFs) designed for platforms with hardware accelerators.

In this proceedings we address the issue of the generation of GPU-ready matrix elements.
We present {\tt MadFlow}, a framework which provides an
automatic pipeline for GPU deployment of Monte Carlo simulation for custom
processes, by combining the matrix elements and phase space expressions
generated by the MadGraph5\_aMC@NLO (MG5\_aMC)~\cite{Alwall:2014hca, Frederix:2018nkq} framework
with the VegasFlow and PDFFlow efficient simulation tool for hardware
accelerators. {\tt MadFlow} will be deployed as a new open-source library,
opening the possibility to study and benchmark multiple approaches to Monte
Carlo integration based on distributed hardware, and, in future, new algorithms
based on deep learning techniques.

This draft is structured as follows. In Section~\ref{sec:methodology} we
describe the technical implementation of our framework. In
Section~\ref{sec:results} we compare and benchmark results. Finally, in
Section~\ref{sec:outlook} we present our conclusion and future development
roadmap.

\section{Technical implementation}
\label{sec:methodology}

\subsection{The {\tt MadFlow} design}
We appreciate that many groups nowadays rely on very extensive code bases
and learning to use an equally complicated framework (no matter the benefit)
might not be feasible for everyone.
Therefore, in order to accelerate the adoption of hardware accelerators
within the HEP-ph community our main concerns when designing a Monte Carlo implementation on GPU
have been the maintainability of the framework and making it as developer-friendly as possible.

In order to achieve our goals, we consider the MG5\_aMC framework~\cite{Alwall:2014hca, Frederix:2018nkq}
as the entry point of our procedure. MG5\_aMC is a meta-code written in Python, that generates
automatically the set of instructions in a low-level language to be employed for

the simulation of arbitrary scattering processes at colliders, in the Standard Model or beyond. MG5\_aMC
relies on general procedures and methods, without being tailored to a specific process, class of processes, or
physics model. Besides the generation of tree-level matrix elements, MG5\_aMC
gives also the possibility to the user to include Next-to-Leading order corrections,
both due to strong and electroweak interactions (including matching to parton-shower for the
former). However, in this paper we will limit ourselves to the case of tree-level matrix elements.

The workflow of MG5\_aMC is the following: a model, written in the Universal
Feynrules Output (UFO) format~\cite{Degrande:2011ua}, is loaded, which contains
all the informations on the underlying theory, including the Feynman rules.
Starting from the model, Feynman diagrams are generated, and the corresponding
expressions for the matrix elements are written in process-specific files. The
various parts of the Feynman diagrams (external wavefunctions, vertices,
propagators, etc.) are evaluated via the ALOHA routines~\cite{deAquino:2011ub}
(with the introduction of MadGraph5~\cite{Alwall:2011uj} ALOHA supersedes the
HELAS routines~\cite{Murayama:1992gi}).

It is then natural to consider a GPU extension of MG5\_aMC (and of of ALOHA) as
the basis of our future general purpose parton-level GPU MC generator. For
consistency with the current string of names, we have dubbed this framework with
the provisional {\tt MadFlow}. We note that there is an independent effort
dedicated to porting MG5\_aMC to GPU~\cite{madgraph4gpu}. The {\tt MadFlow}
project differs from ``Madgraph 4 GPU'' for two main reasons: the interest in
providing the VegasFlow integration engine and thus the possibility to deploy
the algorithm on multiple hardware accelerators, and the need of a technical
solution which simplifies maintainability and does not require specialized GPU
knowledge from the developer and user point of view. Nevertheless, by design
{\tt MadFlow} can import custom CUDA kernels with optimized code from any
external source, thus we might consider interfacing with other projects (such as Madgraph 4 GPU) in the
future.

\begin{figure}
  \centering
  \includegraphics[width=0.95\textwidth]{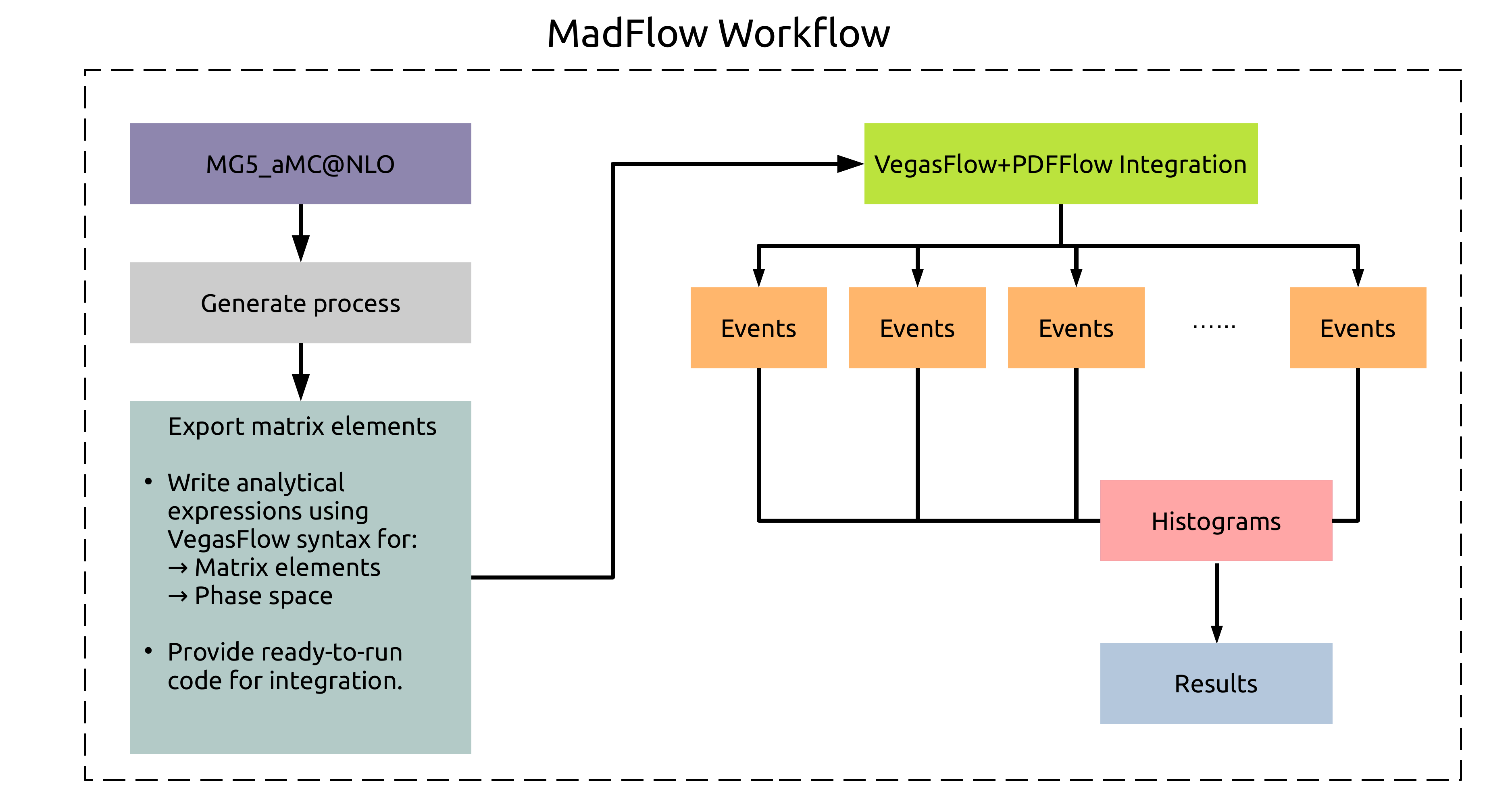}
  \caption{\label{fig:scheme}Schematic workflow for the implementation of {\tt
  MadFlow}. The process starts directly from MG5\_aMC, where the user
  generates a custom process using the standard framework API and exports the
  relevant code in a specific format for VegasFlow and PDFFlow integration.}
\end{figure}

In Figure~\ref{fig:scheme} we show schematically the stages involved in the
implementation of {\tt MadFlow}. The process starts directly from MG5\_aMC,
where the user generates a custom process using the standard framework API and
exports the relevant code for the analytic matrix elements and phase space
expressions in Python, using the syntax defined by VegasFlow. In terms of code
implementation this step requires the development of a MG5\_aMC plugin, which
consists of an exporter module to write the matrix element and the ALOHA
routines in Python, fulfilling the requirements imposed by VegasFlow and PDFFlow using
TensorFlow~\cite{tensorflow2015:whitepaper} primitives. The main difficulty
consists in converting sequential functions into vectorized ones. This has

been achieved both for the matrix element and for the ALOHA routines.
During the {\tt MadFlow} development we have performed several numerical
and performance benchmarks in order to avoid potential bugs.

After the conversion step into the specified format is performed, the exported
Python code is incorporated into a VegasFlow device-agnostic integration
algorithm which executes the event generation pipeline from the generation of
random numbers, computation of the phase space kinematics, matrix element
evaluation and histogram accumulation.

\subsection{The evaluation of matrix elements routines}

The evaluation of the matrix elements in {\tt MadFlow} follows the original
MG5\_aMC implementation: a {\tt Matrix} class is produced by the Python exporter
plugin module. Its {\tt smatrix} method links together the needed Feynman rules
to compute the requested matrix element: it loops over initial and final state
particle helicities and aggregates their contribution to the final squared
amplitude.

The matrix element vectorization requires to replace the ALOHA waveforms and
vertices routines abiding by the TensorFlow ControlFlow rules. Although
this process is straightforward for vertices Feynman rules, being mostly comprised
by elementary algebraic operations, the implementation of particle waveforms
functions is subject to several conditional control statements that definitely make the
task harder.

\section{Preliminary results}
\label{sec:results}

We have tested the output of {\tt MadFlow} 
in a Leading Order calculation for gluon-induced top-pair production at $\sqrt{s}=14$ TeV.
Note that, in principle, any process generated (at LO) by the MG5\_aMC framework can be
integrated by the {\tt MadFlow} library.

\subsection{Accuracy}

\begin{figure}
  \centering
  \includegraphics[width=0.9\textwidth]{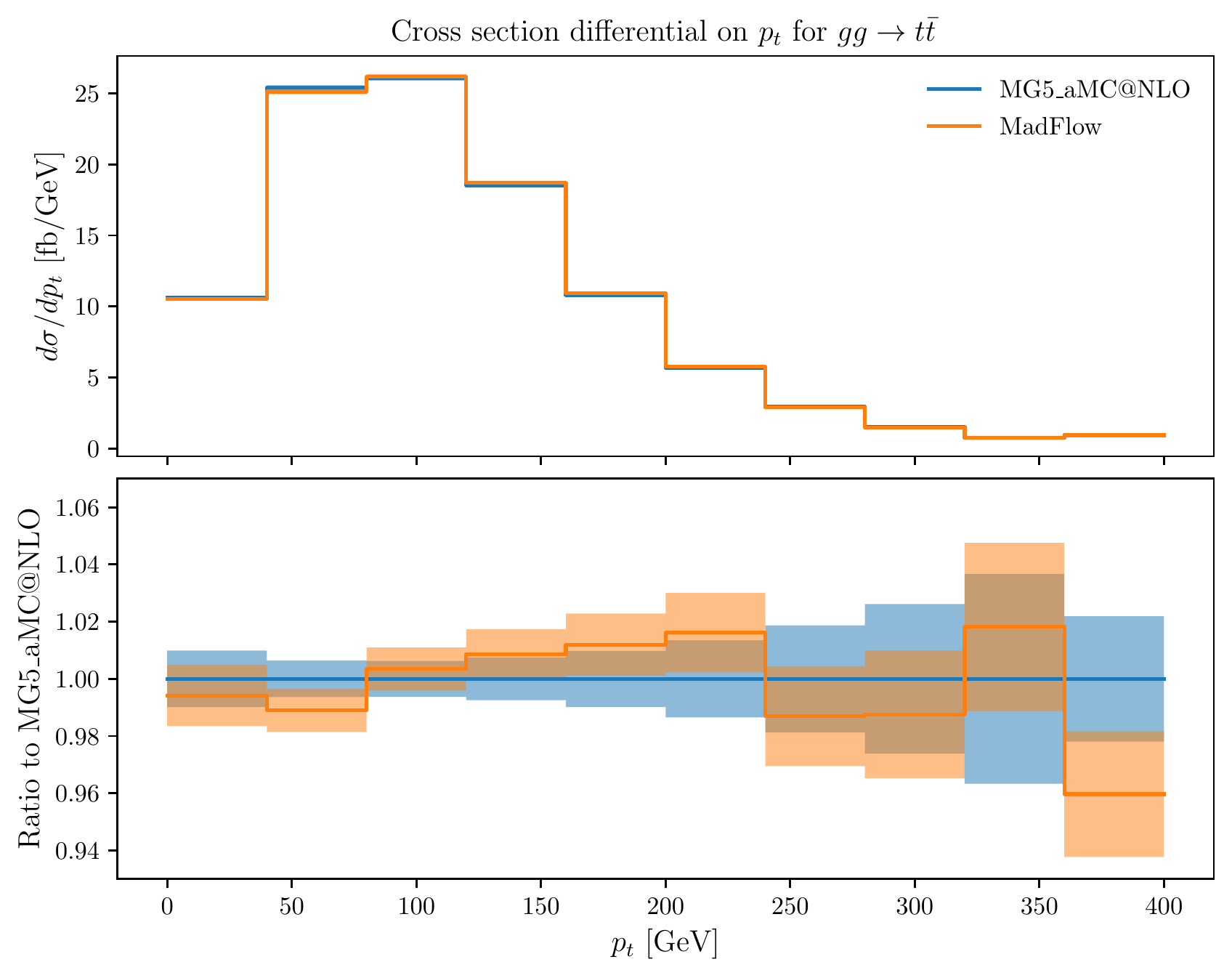}
  \caption{\label{fig:ggpt}Leading Order cross section differential on $p_t$ for
  gluon-gluon fusion at $\sqrt{s}=14$ TeV. We compare predictions between
  MG5\_aMC (blue) and VegasFlow (orange), using the conversion tool
  described in Section~\ref{sec:methodology}. The two cross sections are in statistical agreement.}
\end{figure}

In Figure~\ref{fig:ggpt} we show the Leading Order cross section differential on
$p_t$ for gluon-gluon fusion for predictions obtained with the original MG5\_aMC
integration procedure and the new {\tt MadFlow} approach based on VegasFlow and
PDFFlow. The plot in the first row shows the differential distribution in
absolute scale in fb/GeV, while the plot in the second row show the ratio
between both computations, confirming a good level of agreement between both
implementations for the same level of target accuracy of $0.02\%$.

The results presented here are computed independently from each framework in
order to minimize communication bottlenecks between CPU-GPU, in fact for this
particular configuration {\tt MadFlow} does not require any specific
asynchronous CPU evaluation, thus all computations are performed directly on GPU
and results stored on GPU memory.

\subsection{Performance}

\begin{figure}
  \centering
  \includegraphics[width=0.9\textwidth]{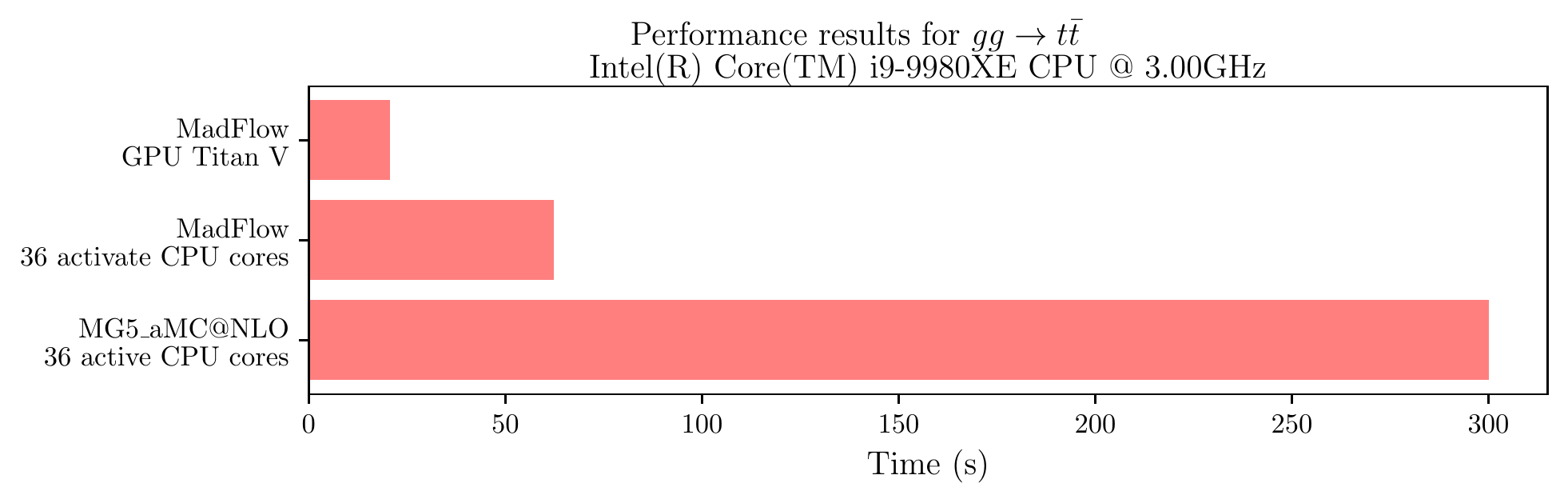}
  \caption{\label{fig:performance}Comparison of a Leading Order cross section
  differential on $p_t$ for gluon-gluon fusion at $\sqrt{s}=14$ TeV ran in both
  VegasFlow and MG5\_aMC. The CPU-only version of VegasFlow is able to
  improve the performance obtained by MG5\_aMC for the same level of target
  accuracy due mainly to a difference on the parallelization implementation of both codes.
  The usage of the GPU device further improves the performance.}
\end{figure}

In terms of performance, in particular evaluation time, in
Figure~\ref{fig:performance} we compare the total amount of time required by the
simulation using three different configurations based on a workstation with an
Intel i9-9980XE CPU with 36 logical cores. The first horizontal bar shows the
greatest performance of {\tt MadFlow} running on a single NVIDIA Titan V GPU
with 12GB of RAM. The second bar shows the timings for the same code evaluated
just on the CPU chip using all available cores. The last bar shows the equivalent
simulation performance using the MG5\_aMC default integrator.
We conclude that the {\tt MadFlow} implementation confirms a great performance
improvement when running on GPU hardware.

\section{Outlook}
\label{sec:outlook}

In conclusion in this proceedings we propose a new approach for the
generalization of Monte Carlo simulation on hardware accelerators. In
particular, the {\tt MadFlow} design provides a fast and maintainable code which
can quickly port complex analytic expressions into hardware specific languages
without complex operations involving several computer languages, tools and
compilers.
Furthermore, preliminary results confirm the algorithm effectiveness, providing
benefits in terms of performance when compared to CPU based simulations.

As an outlook, we plan to release a development version of {\tt MadFlow} as an
open-source library, and proceed with the implementation of all required changes
in order to accommodate Next-to-Leading order computations in the future.

\section*{Acknowledgements}

We thank Olivier Mattelaer, Andrea Valassi and Stefan Roiser for discussions
about GPU kernel optimization for matrix elements. S.C. and J.C.M are supported
by the European Research Council under the European Union's Horizon 2020
research and innovation Programme, grant agreement number 740006, valid until
the end of 2021, exclusively for research topics related to the NNNPDF project.
M.Z.~is supported by Programma per Giovani Ricercatori ``Rita Levi Montalcini''
granted by the Italian Ministero dell'Universit\`a e della Ricerca (MUR).

%
\bibliography{blbl.bib}

\end{document}